\documentclass[8.5pt,preprint]{article}

\usepackage[super,sort&compress,comma]{natbib} 
\usepackage{mhchem}
\usepackage{times,mathptmx}
\usepackage{sectsty}
\usepackage{balance} 

\usepackage{graphicx} 
\usepackage{lastpage}
\usepackage[format=plain,justification=raggedright,singlelinecheck=false,font=small,labelfont=bf,labelsep=space]{caption} 
\usepackage{fancyhdr}
\usepackage{color}
\definecolor{red}{rgb}{1,0.1,0.1} 
\definecolor{blue}{rgb}{0.1,0.1,1} 
\definecolor{verd}{rgb}{0,1,0} 

\begin{document}

\thispagestyle{plain}
\fancypagestyle{plain}{
\renewcommand{\headrulewidth}{1pt}}
\renewcommand{\thefootnote}{\fnsymbol{footnote}}
\renewcommand\footnoterule{\vspace*{1pt}%
\hrule width 3.4in height 0.4pt \vspace*{5pt}} 
\setcounter{secnumdepth}{5}

\makeatletter 
\def\subsubsection{\@startsection{subsubsection}{3}{10pt}{-1.25ex plus -1ex minus -.1ex}{0ex plus 0ex}{\normalsize\bf}} 
\def\paragraph{\@startsection{paragraph}{4}{10pt}{-1.25ex plus -1ex minus -.1ex}{0ex plus 0ex}{\normalsize\textit}} 
\renewcommand\@biblabel[1]{#1}            
\renewcommand\@makefntext[1]%
{\noindent\makebox[0pt][r]{\@thefnmark\,}#1}
\makeatother 
\renewcommand{\figurename}{\small{Fig.}~}
\sectionfont{\large}
\subsectionfont{\normalsize} 

\fancyfoot{}
\fancyfoot[RO]{\footnotesize{\sffamily{1--\pageref{LastPage} ~\textbar  \hspace{2pt}\thepage}}}
\fancyfoot[LE]{\footnotesize{\sffamily{\thepage~\textbar\hspace{3.45cm} 1--\pageref{LastPage}}}}
\fancyhead{}
\renewcommand{\headrulewidth}{1pt} 
\renewcommand{\footrulewidth}{1pt}
\setlength{\arrayrulewidth}{1pt}
\setlength{\columnsep}{6.5mm}
\setlength\bibsep{1pt}

\twocolumn[
  \begin{@twocolumnfalse}
\noindent\LARGE{\textbf{Effective Attractive Range and Viscoelasticity of Colloidal Gels}}
\vspace{0.6cm}

\noindent\large{\textbf{P. H. S. Santos,\textit{$^{a}$} O. H. Campanella,\textit{$^{a}$} and
M. A. Carignano$^{\ast}$\textit{$^{b}$}}}\vspace{0.5cm}

\noindent\textit{\small{\textbf{Received Xth XXXXXXXXXX 20XX, Accepted Xth XXXXXXXXX 20XX\newline
First published on the web Xth XXXXXXXXXX 200X}}}

\noindent \textbf{\small{DOI: 10.1039/b000000x}}

\noindent \normalsize{We present a simulation study of colloidal particles having very short range attractions. In particular, we investigate the interplay between the effective attractive range and the viscoelastic properties of the gels that form when the temperature of the system is sufficiently low. Using Brownian dynamics simulations we study particles with different size and interaction range, and characterize the mechanical properties of the gels by performing small amplitude oscillatory and transient starts-up shear simulations. We found that the effective attractive range is a key parameter affecting the gel viscoelastic properties. The crossover frequency between the storage and loss moduli and the time to reach the maximum stress in the start-up test are both simple functions of the effective attractive range.}
\vspace{0.5cm}
 \end{@twocolumnfalse}
  ]

\section{Introduction}


\footnotetext{\textit{$^{a}$~Department of Agricultural and Biological Engineering, Purdue University, 225 South University Street, West Lafayette, IN 47907, USA.}}
\footnotetext{\textit{$^{b}$~Qatar Environment and Energy Research Institute, P.O. Box 5825, Doha, Qatar. E-mail=mcarignano@qf.org.qa}}



The addition of {\em gelling} colloidal particles to a low molecular weight solvent produces a drastic effect on the flowing characteristics of the solvent. The result of the mix is a material with a solid like appearance, in the sense that it does not flow unless under the effect of a sufficiently large external stress.\cite{FLORY:1975aa,Terech:1997aa,Dietsch:2008aa} Such type of soft materials find applications in a wide range of technological uses. Nanocomposite gels, for instance, emerge as an alternative in tissue engineering, cell culture, drug delivery systems and biomedical devices.
\cite{Guvendiren:2012aa,Al-Ghanami:2010aa,Wang:2011aa,Freemont:2008aa,Pek:2008aa,Wang:2010aa,Xie:2006aa} Colloidal particles are also the basis of a number of food and personal products, key components in the coating industry and have been investigated as an alternative to improve safety and efficiency in rocket propulsion systems.\cite{Dickinson:2012aa,Le-Reverend:2010aa,Dickinson:2006aa,Arnold:2011aa,Madlener:2012aa}

Colloidal gels can be thought as a space filling or percolating network of particles. \cite{Zaccarelli:2007aa,Lu:2008aa,Schooneveld:2009aa,Ruzicka:2011aa} Particulate suspensions can form a percolating network when the system is destabilized. Under appropriate thermodynamic conditions, sufficient particle concentration and provided that the attraction between the particles are strong enough to induce aggregation, a space filling macroscopic structure is formed that effectively trap the solvent molecules.\cite{Rajaram:2010aa,Mezzenga:2005aa,Zaccarelli:2005aa,Fernandez-Toledano:2009aa} Although not a thermodynamical equilibrium state, the percolating network undergoes a dramatic kinetic slowdown and the resulting gel has a long life.\cite{Foffi:2005aa,Lodge:1999ab,Lodge:1998aa,LIN:1989aa,Sorensen:2011aa} The main parameters that affect this kinetic slowdown are the range of the attractive potential and its relation to the particle size, and the particle density.\cite{Lodge:1999aa,Trappe:2004aa} It is also important whether the contact of two particles results in a permanent bond, or if the particles have the additional freedom of rotating  on top of each other. In the first case, the formation of the network is dominated by the diffusion\cite{LIN:1989aa} while in the second case there is a slow kinetic evolution driven by the phase separation occurring in the system.\cite{Kamp:2009aa,Trappe:2001aa} We focus our attention on the latter case, where the inter-particle interactions is described by spherical, pair-wise additive potentials.

{ The range of the inter-particle interaction has an important effect on the phase diagram of the system.\cite{TEJERO:1994aa} This was studied for many model systems, like $2n-n$ Lennard-Jones\cite{Vliegenthart:1999aa} and hard-core Yukawa potentials\cite{Foffi:2002aa}. The general result is that as the range of the attractive tail is reduced, the gas-liquid coexistence becomes limited to smaller density region. For short enough attractive range, the liquid phase completely disappears, leaving only the possibility of a coexistence between a gas and solid phases.\cite{Hasegawa:1998aa}} From an experimental point of view, it has been clearly established that long range attractions lead to the collapse of the gel in relatively short times.\cite{Teece:2011aa,Bartlett:2012aa} Although our understating of the relationship between the range of attractive potential and shape of phase diagram has been improved along the years, insights about how the potential's attractive range affects the mechanical properties of the resulting colloidal gels are still lacking. The difficulties to experimentally control the wide variety of parameters involved in these systems make computational studies a useful tool to investigate the role of these parameters in order  to design optimal gels for specific applications. Most theoretical studies have been focused on the gelation routes and have been trying to develop a general theoretical framework covering a broad range of particle concentrations.\cite{Zaccarelli:2007aa} The present work focuses on the role of the potential range on the viscoelastic properties of colloidal gels. Particulate systems interacting via $R$-shifted 12-6 Lennard-Jones potential were simulated using Brownian Dynamics to study the structural responses of colloidal gels to small and large deformations. Our results show that the attractive range plays an important role in the viscoelastic behavior of particle gels, and also suggest how a dimensionless interaction range could be determined using start-up shear experiments.

\section{Simulation details and Theoretical Background}

{ We have studied the formation and properties of gels using Brownian dynamics simulations in the over damped limit. \,\, Namely, we numerically integrate the position Langevin equation:
\begin{equation}
\frac{dr_i}{dt}=\frac{1}{\gamma} f_i(t) + \theta_i(t) \label{bdeq}
\end{equation}
where $\gamma$ is the friction constant, $f_i(t)$ is the force acting on particle $i$ due to the interaction with the other colloidal particles and $\theta_i(t)$ is a random noise satisfying $\langle \theta_i(t) \theta_j(0) \rangle = 2 \delta_{ij} \delta(t) k_BT/\gamma$. }
The interaction between particles was described by a $R$-shifted Lennard-Jones potential, as in our previous study:\cite{Santos:2010aa}
\begin{equation}
	U(r) = \left\{ 
	\begin{array}{ll}
 	\infty &\mbox{, if $r \leq R_0$} \\
  	4 \epsilon \left [ \left ( \frac{\sigma}{r-R_0} \right )^{12} - \left ( \frac{\sigma}{r-R_0} \right )^{6} \right ]  & \mbox{, if $r > R_0$}
       	\end{array} 
	\right.
\end{equation}
where $\epsilon$ is the minimum of the potential that takes place at $r=R_0+2^{1/6}\sigma$. { This potential allows for an independent control of the the size of the particle, approximately equal to $R_0+\sigma$, and the effective range of the attractive tail, controlled by $\sigma/R_0$.} We performed a series of simulations using different combinations of $R_0$ and $\sigma$, as described in Table \ref{tabla}. For all the cases we set $\epsilon=1$. For convenience we define  the reduced temperature $T^*=k_B T/\epsilon$, where $k_B$ is the Boltzman constant. All the simulations were performed using $N=2000$ particles, and run under constant volume conditions. The reduced temperature $T^*$ regulates the random noise acting over the particles.

\begin{table}[h]
\small
\caption{Summary of the different simulated systems.}
\label{tabla}
  \begin{tabular*}{0.47\textwidth}{@{\extracolsep{\fill}}rcc | rcl | l}
\hline
$R_0$ & $\sigma$ & $T^*$ & $\sigma_e$ & $\phi$ & $R_e$ & Label\\
\hline
  9.5 & 0.5 & 0.20 & 10.032 & 0.1313 & 0.0088 & A1\\
  9.0 & 1.0 & 0.20 & 10.064 & 0.1325 & 0.0179 &  A2 \\ 
  4.0 & 1.0 & 0.20 &   5.064 & 0.1351 & 0.0353 & A3\\
  8.0 & 2.0 & 0.20 &  9.122 & 0.0987 & 0.04697 & A4\\ 
19.0 & 1.0 & 0.12 & 20.075 & 0.1315 & 0.0063 & B1\\
  9.0 & 1.0 & 0.12 & 10.075 & 0.1329 & 0.0127 & B2\\
  9.0 & 1.0 & 0.08 & 10.083 & 0.1332 & 0.00987 & C\\
\hline
\end{tabular*}
\end{table}

In Brownian dynamics simulations, it is customary and convenient to define the unit of time as the time needed by an isolated particle to diffuse a distance proportional to its size. Therefore, since we are dealing with a soft spherical potential, we need to introduce a uniform definition for the particle size. Following Weeks-Chandler-Andersen theory\cite{WEEKS:1971aa}, we define $\sigma_e$, the effective diameter of the particles as:
\begin{equation}
	\sigma_{e}=\int_0^\infty \left [ 1 - e^{\beta U_{r}(r)} \right ] dr
\end{equation}
where $U_r(r)$ is the repulsive part of the interaction potential and $\beta=1/k_BT$. Assuming that the particle diameter is $\sigma_e$ we define the time unit as  $\tau=\sigma_e^2/4D_0$, where  $D_0$ is the diffusion coefficient of the particles at infinite dilution. { Using this unit, the time step that we employed to integrate Eq. (\ref{bdeq}) is 10$^{-6} \tau$, which is sufficiently small to prevent the particles to reach the region of infinite potential.}The volume fraction of the system is also expressed in terms of $\sigma_e$ as  $\phi=(N/V) \pi \sigma_e^3/6$, with $N$ being the number of particles simulated in a volume $V$.

As it will become clear below, it is also convenient to introduce the concept of effective range of attraction. The range of attraction of an arbitrary interaction potential is a vague concept, and difficult to quantify when dealing with continuous attractive tails. On the contrary, the square well potential has a univocally defined range of attraction. Let us consider a square well potential defined by:
\begin{equation}
U_{sw}(r) = \left\{ \begin{array}{ll}
 \infty &\mbox{, if $r \leq \sigma$} \\
  -\epsilon  & \mbox{, if $\sigma < r \leq \lambda \sigma $} \label{sw}\\
  0 & \mbox{, if $\lambda \sigma < r$}
       \end{array} \right.
\end{equation}
The dimensionless range of the interaction is defined as the range measured in units of the particle size. 
Then, the dimensionless range for the square well of Eq. (\ref{sw}) is $R=\lambda-1$. Following the work of Noro and Frenkel on extended corresponding states,\cite{Noro:2000aa}  the definition of the effective range of an interaction potential of arbitrary shape is done by establishing a proper connexion to the square well potential. The link between the square well potential and an arbitrary attractive potential is established through the reduced second virial coefficient:
\begin{equation}
	B_2^*=\frac{3}{\sigma_e^3} \int_0^\infty  \left ( 1 - e^{-\beta U(r)} \right ) r^2 dr
\end{equation}
The effective range $R_e$ of an arbitrary attractive potential is then defined as the range of that square well potential that yields the same reduced second virial coefficient at the same reduced temperature. Table \ref{tabla} summarizes the effective ranges for all the simulated conditions. All the simulated systems are characterized by a short effective attractive range, which result in gels having relatively long lives.\cite{Lodge:1999ab,Santos:2010aa,Teece:2011aa}
Moreover, as suggested in our previous work\cite{Santos:2010aa} and several theoretical studies\cite{TEJERO:1994aa,Vliegenthart:1999aa,Foffi:2002aa,Hasegawa:1998aa} our new results confirm that at the simulated conditions the systems are trapped in a liquid-solid separation process.

\section{Results}

\begin{figure}[h]
\centering
\includegraphics[width=7.0cm]{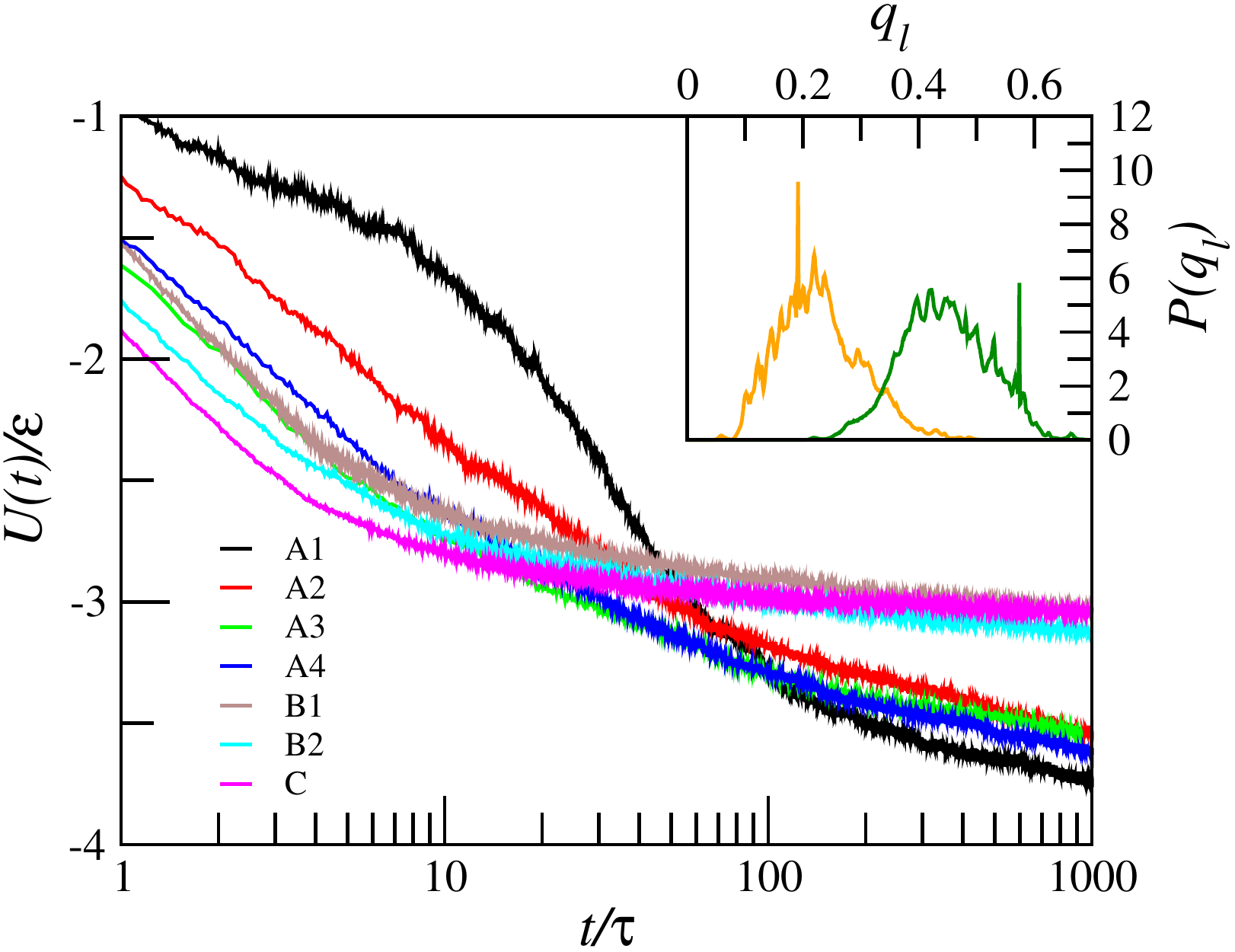}
\caption{Time evolution of the potential energy per particle for all the simulated systems. The different colors correspond to the different systems, as defined in Table \ref{tabla}.  The inset shows the distribution of the order parameters $q_4$ (orange) and $q_6$ (dark green), for the final conformation of system A3.}
\label{epot}
\end{figure}

All the simulations were started using random, high energy, initial conformations with almost no contact between particles. These initial structures were obtained by running a short simulation at high temperature. The system was then destabilized by instantaneous quenching to the target temperature. For all the simulated systems (see Table \ref{tabla}) we observed the formation of a percolated cluster that spanned over the whole simulation box in a short time; approximately $t/\tau \simeq 5$. This aggregation process is reflected by the rapid decrease of the potential energy, as shown in Figure \ref{epot}. As the time progresses, there is a kinetic slow down related to small structural rearrangements in the system. The ordering of the curves in Figure \ref{epot} is mainly dominated by the temperature of the system, but affected to a lesser degree by the effective range of the interactions and the other system parameters. The initial decay of the potential energy reaches lower energies (i.e. higher degree of aggregation) for the systems at lower temperature. However at longer times the opposite behavior is observed. This crossover can be understood in terms of the kinetic slow down of the system, which is strongly dependent on temperature. The higher the temperature, the higher the ability of the system to evolve toward its preferred energy state.  In all the cases, the final state of the system consists of a collection fused flocs, some of them having a crystalline structure. As an example, in the insert of Figure 1 we show distribution of the local order parameters $q_4$ and $q_6$, averaged over a period of 10 $\tau$ at the end of the simulation. The results show peaks at 0.191 ($q_4$) and 0.575 ($q_6$), characteristics of the {\em fcc} structure.\cite{STEINHARDT:1983aa}

\begin{figure}[!t]
\begin{center}
\includegraphics[width=7.0cm]{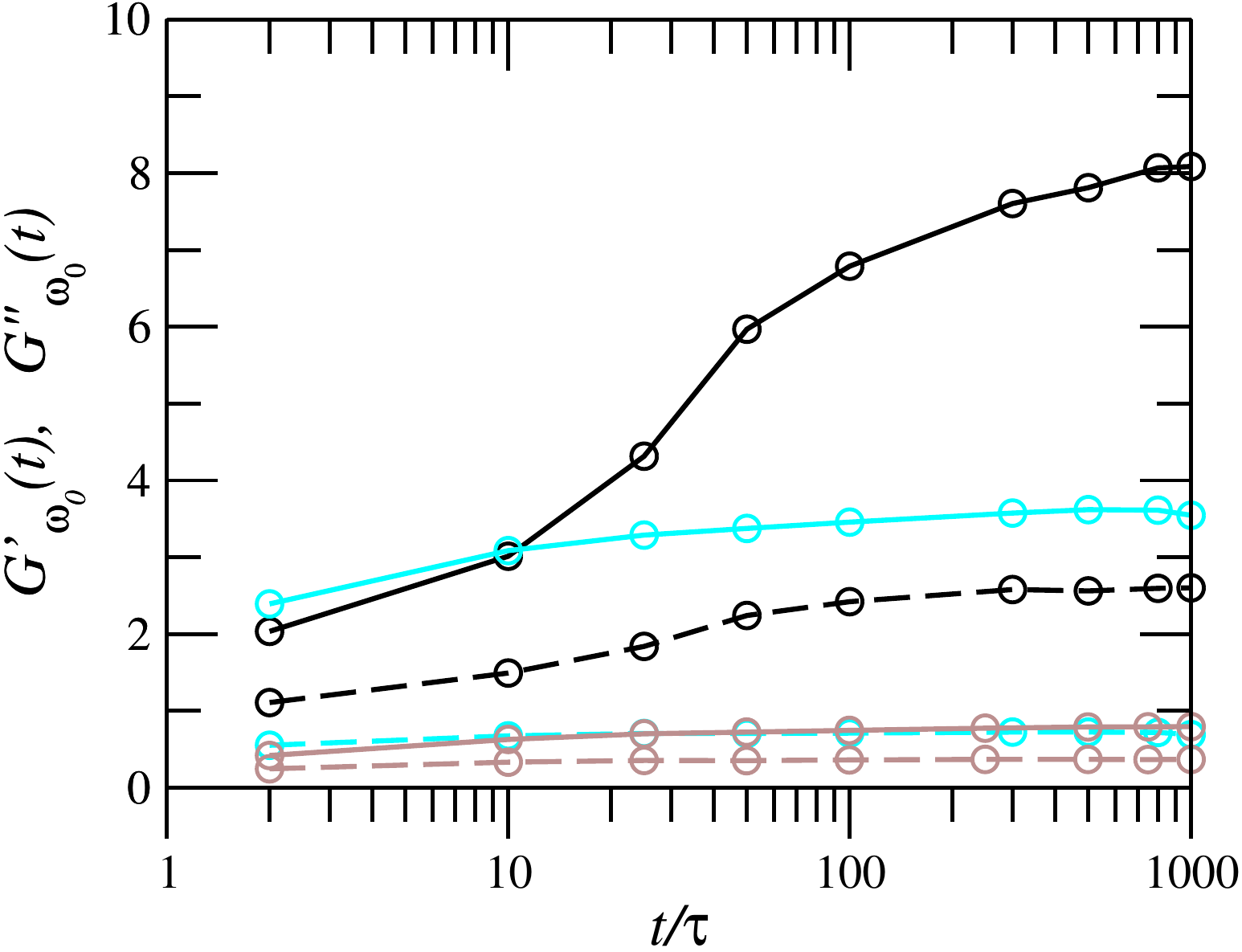}
\end{center}
\caption{Storage (solid lines) and loss (dashed lines) moduli for systems A1, B1 and B2, at a fixed frequency $\omega_0=10^5 \tau^{-1}$ and as a function of time during the gelation process. Colors as in Fig. \ref{epot}}
\end{figure}

\begin{figure}[!b]
\begin{center}
\includegraphics[width=7.0cm]{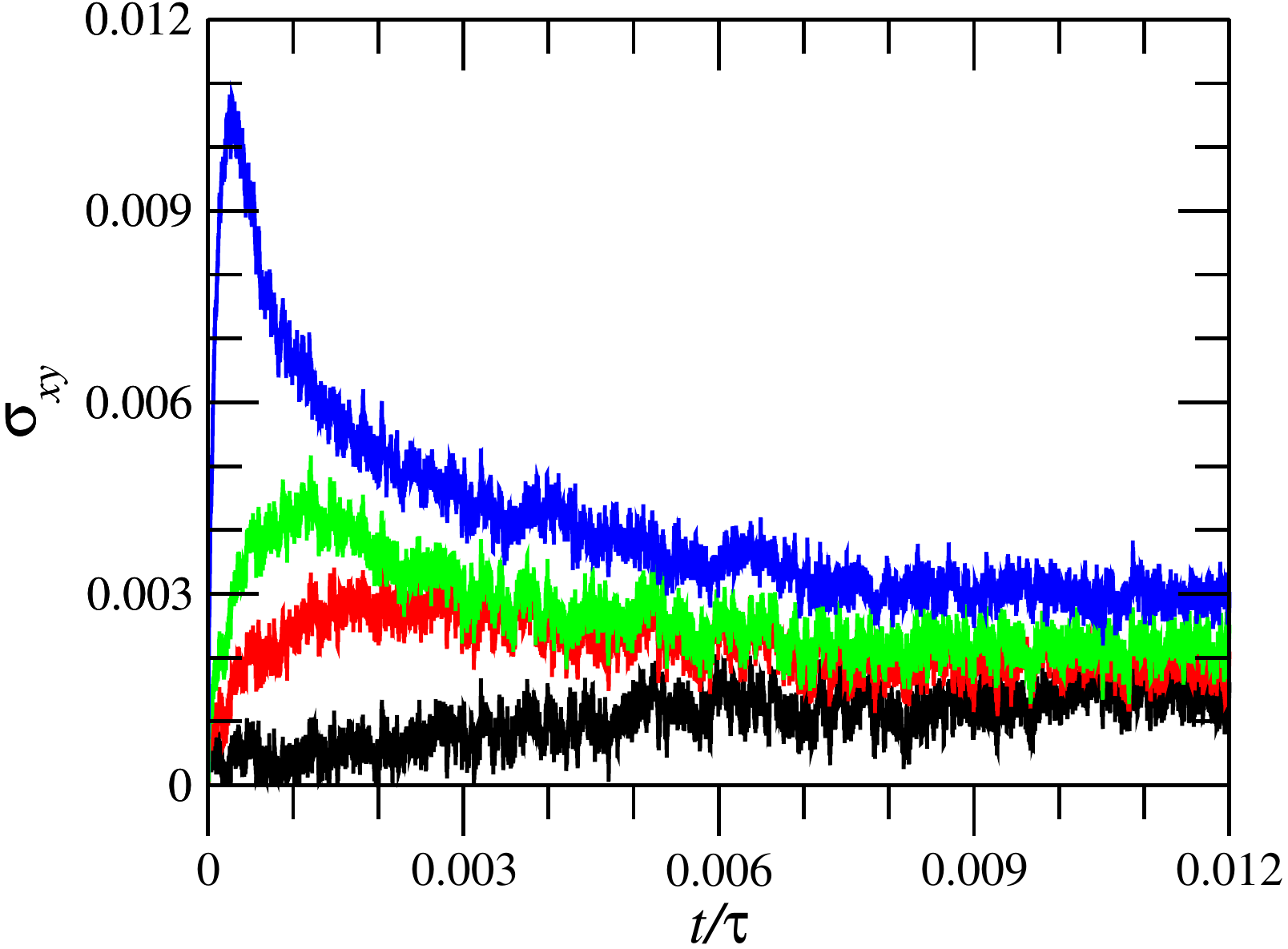}
\end{center}
\caption{Stress overshoot for system B1, for different shear rates: 2$\tau^{-1}$ (black), 10$\tau^{-1}$ (red), 20$\tau^{-1}$ (green) and 100$\tau^{-1}$ (blue)}
\end{figure}

\begin{figure*}[!t]
\begin{center}
\includegraphics*[width=0.3\textwidth]{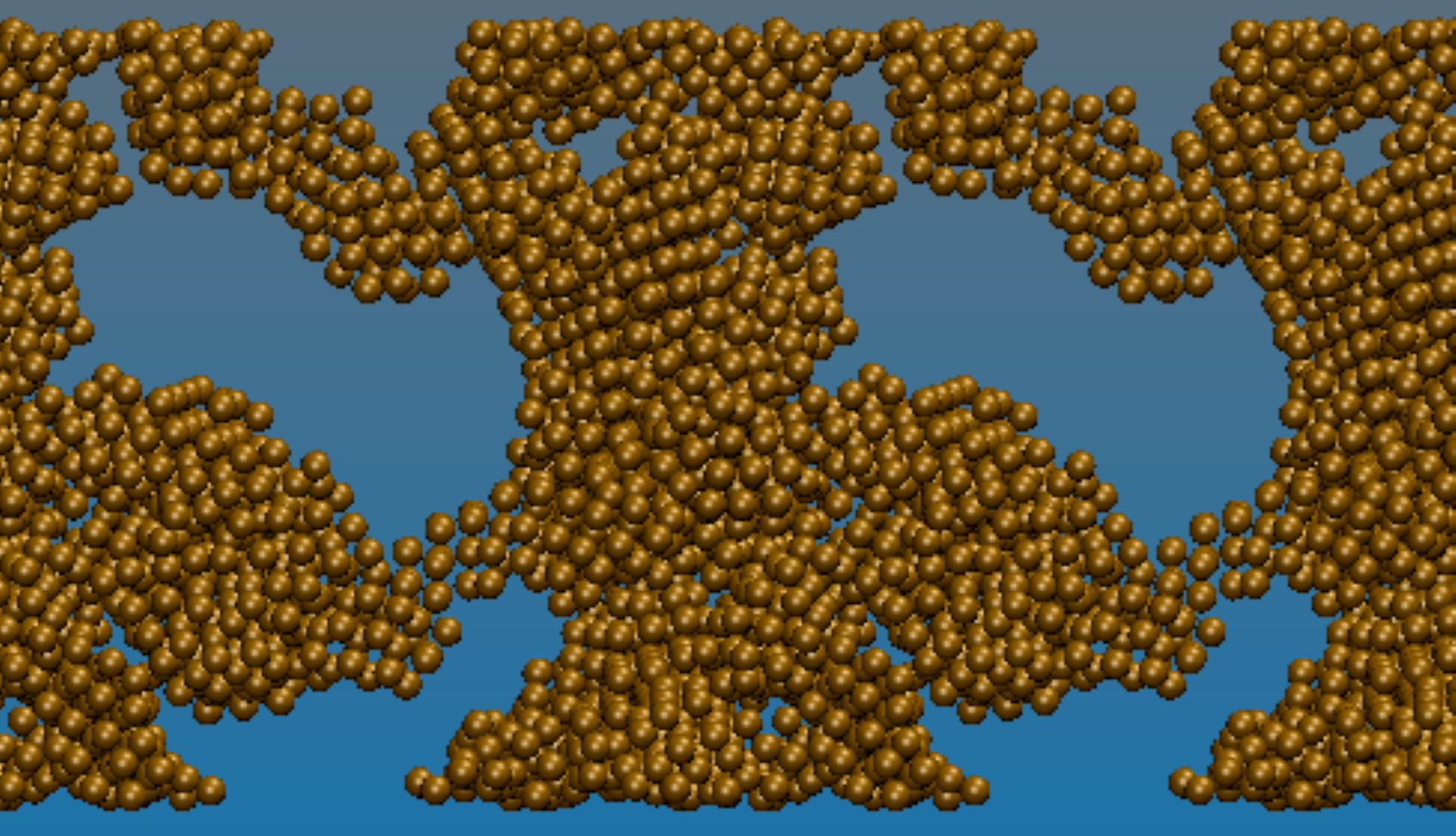} \includegraphics*[width=0.3\textwidth]{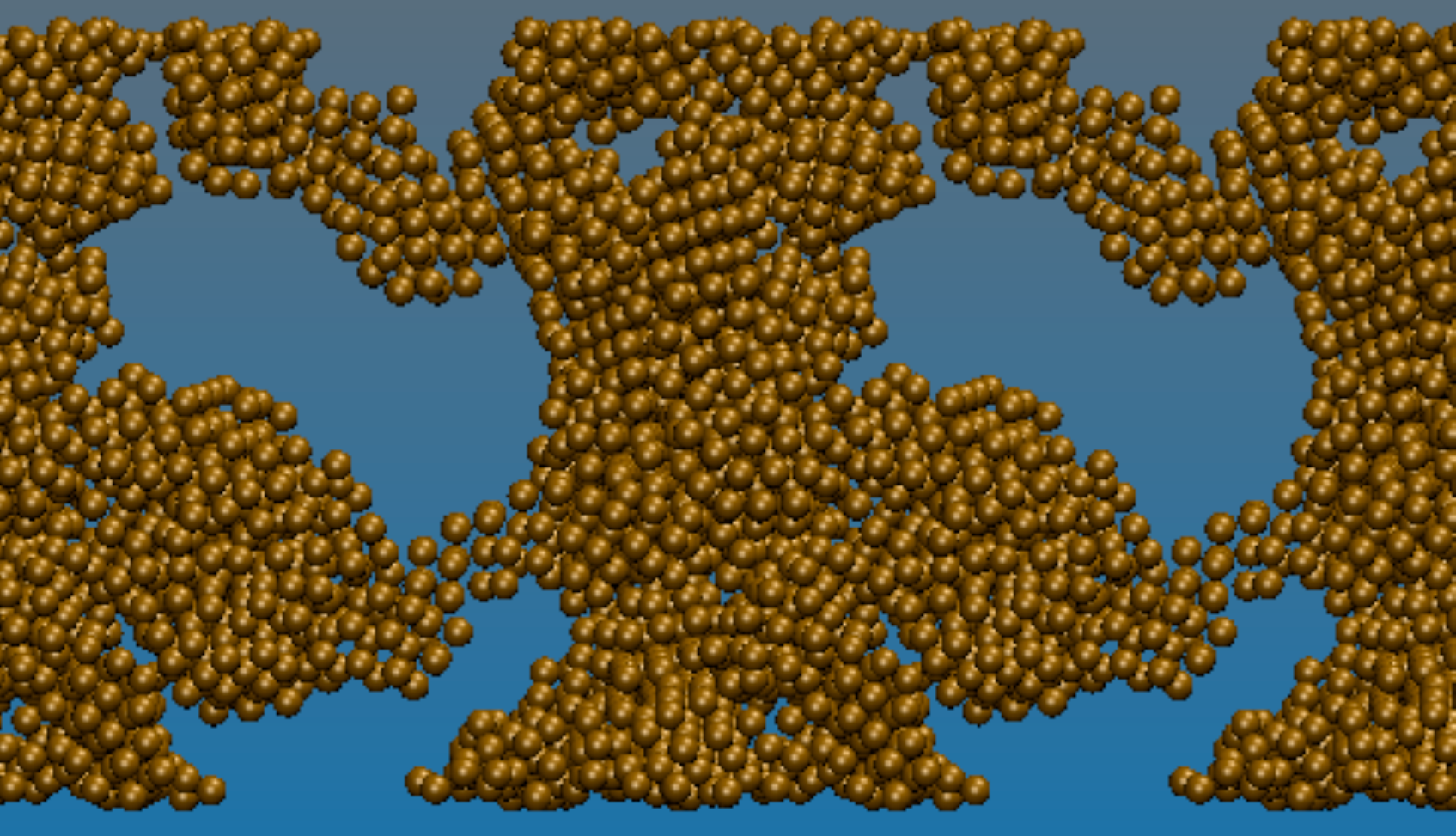} \includegraphics*[width=0.3\textwidth]{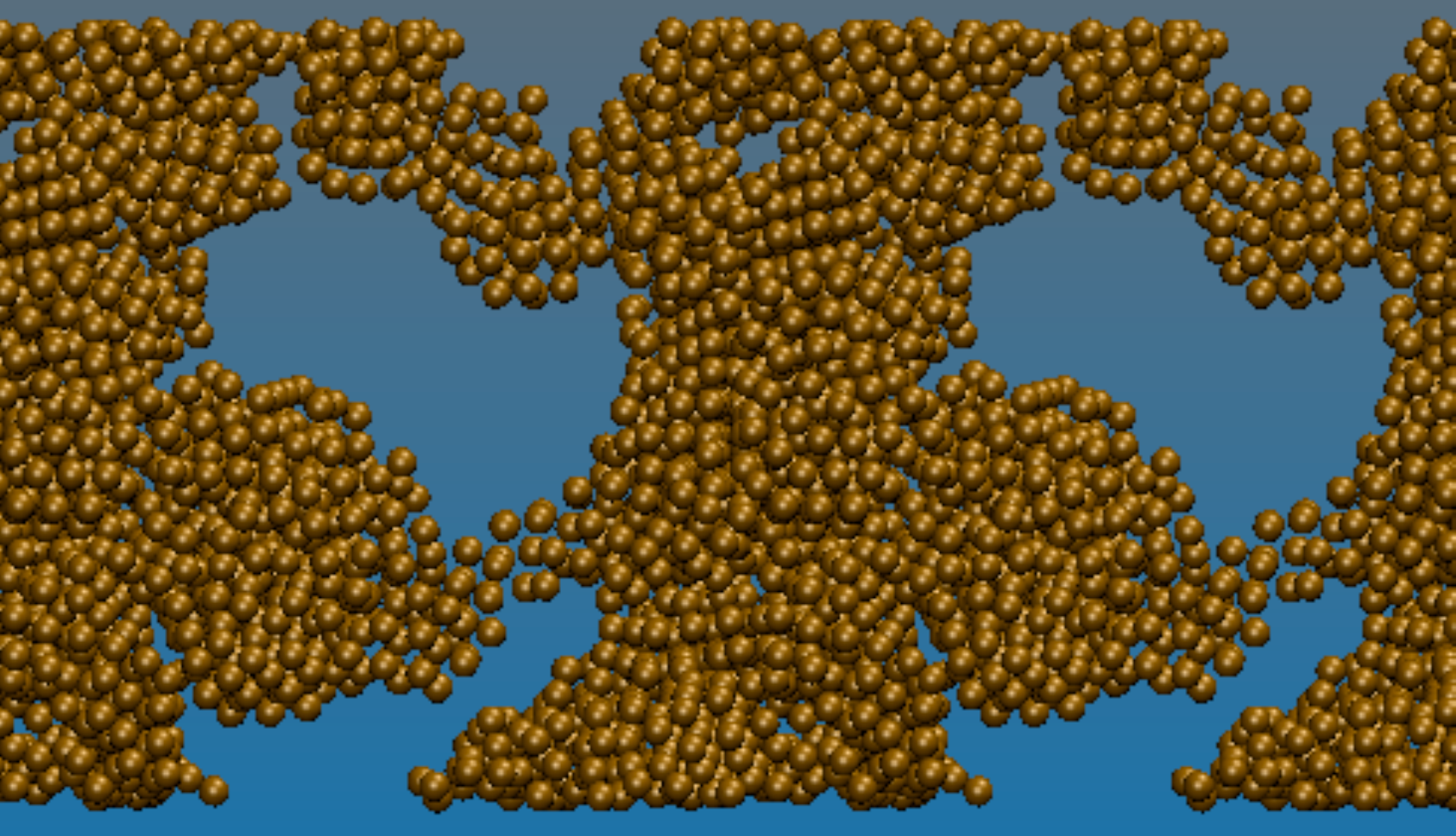}\\
\includegraphics*[width=0.3\textwidth]{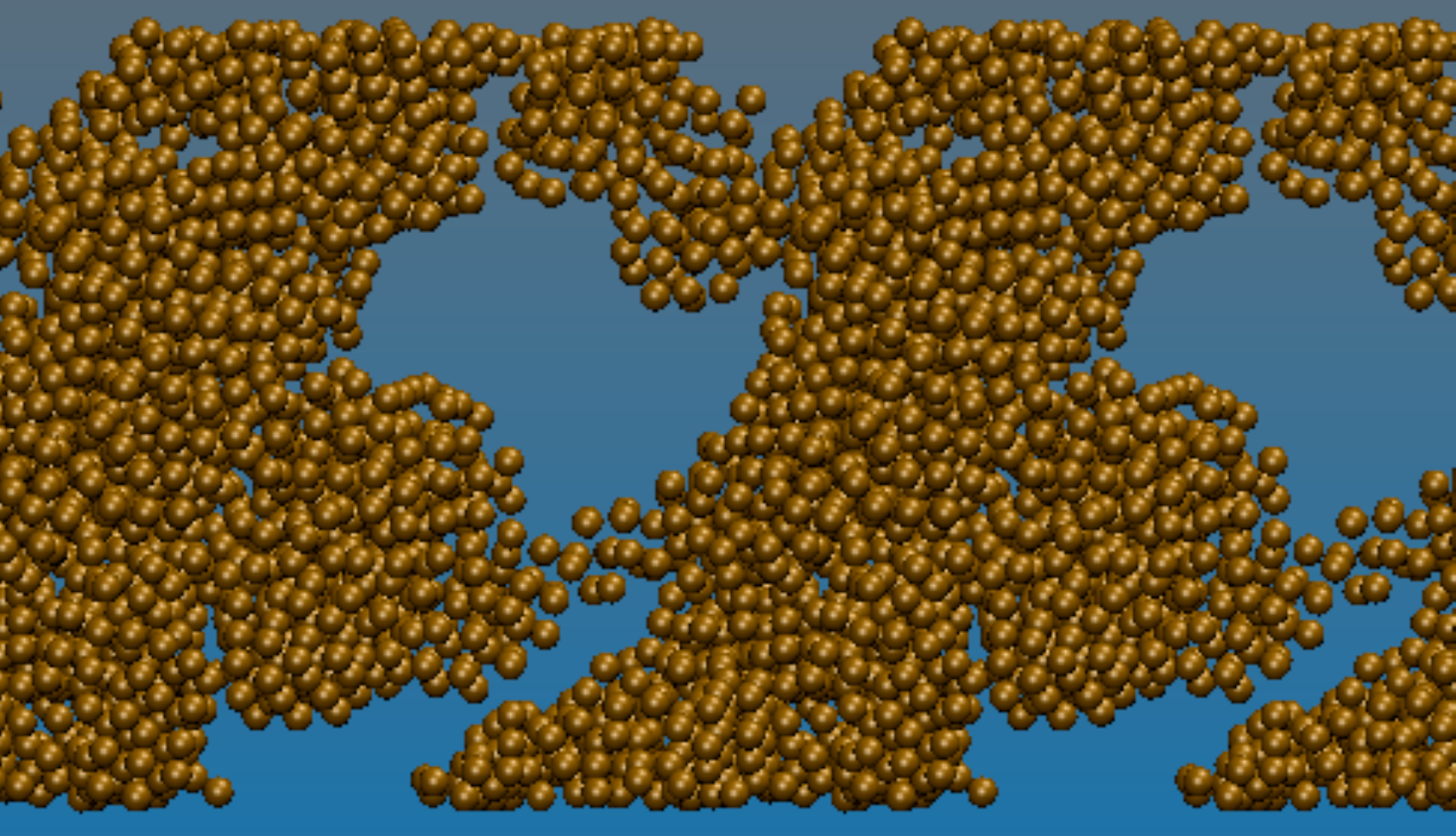} \includegraphics*[width=0.3\textwidth]{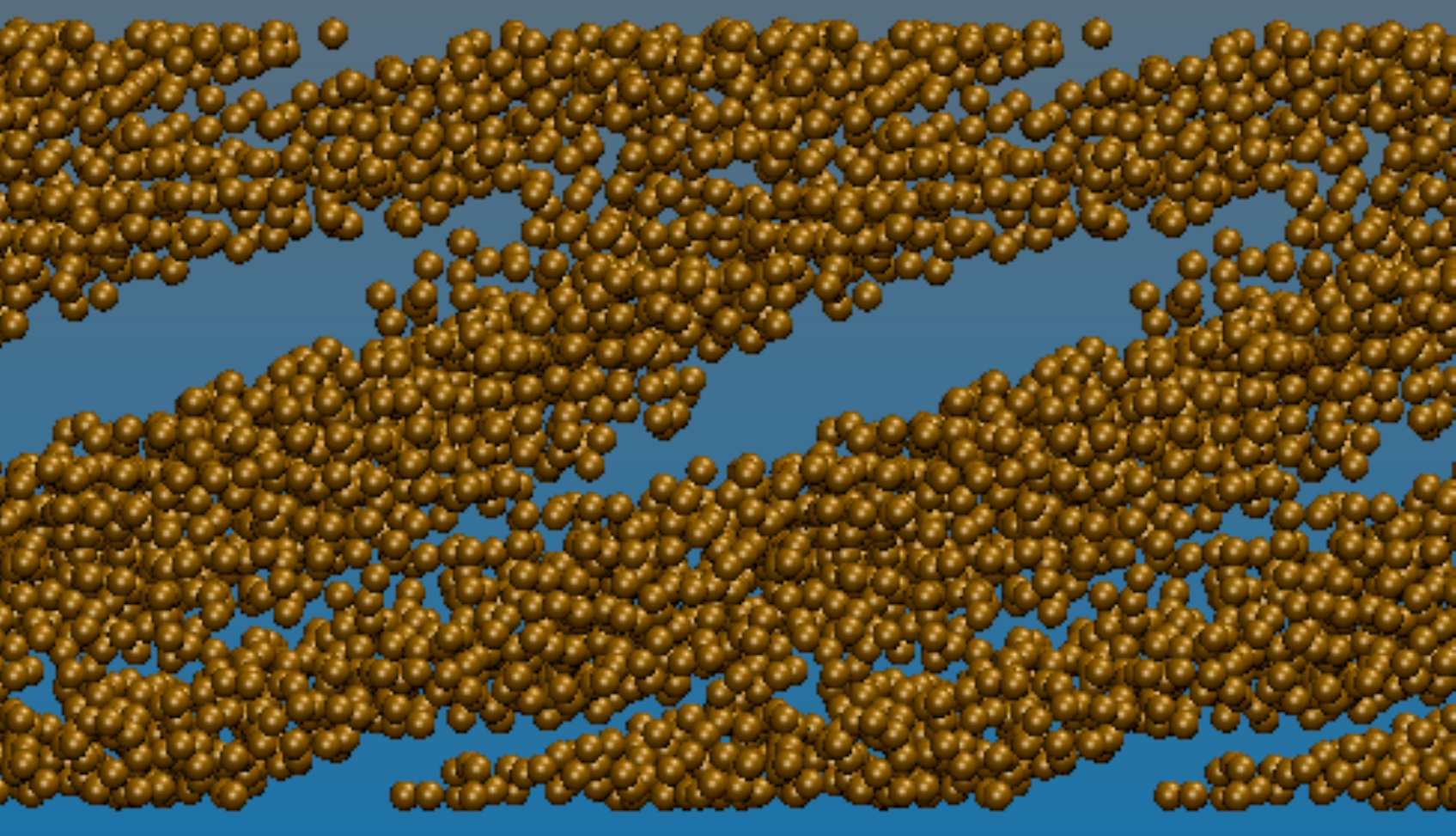} \includegraphics*[width=0.3\textwidth]{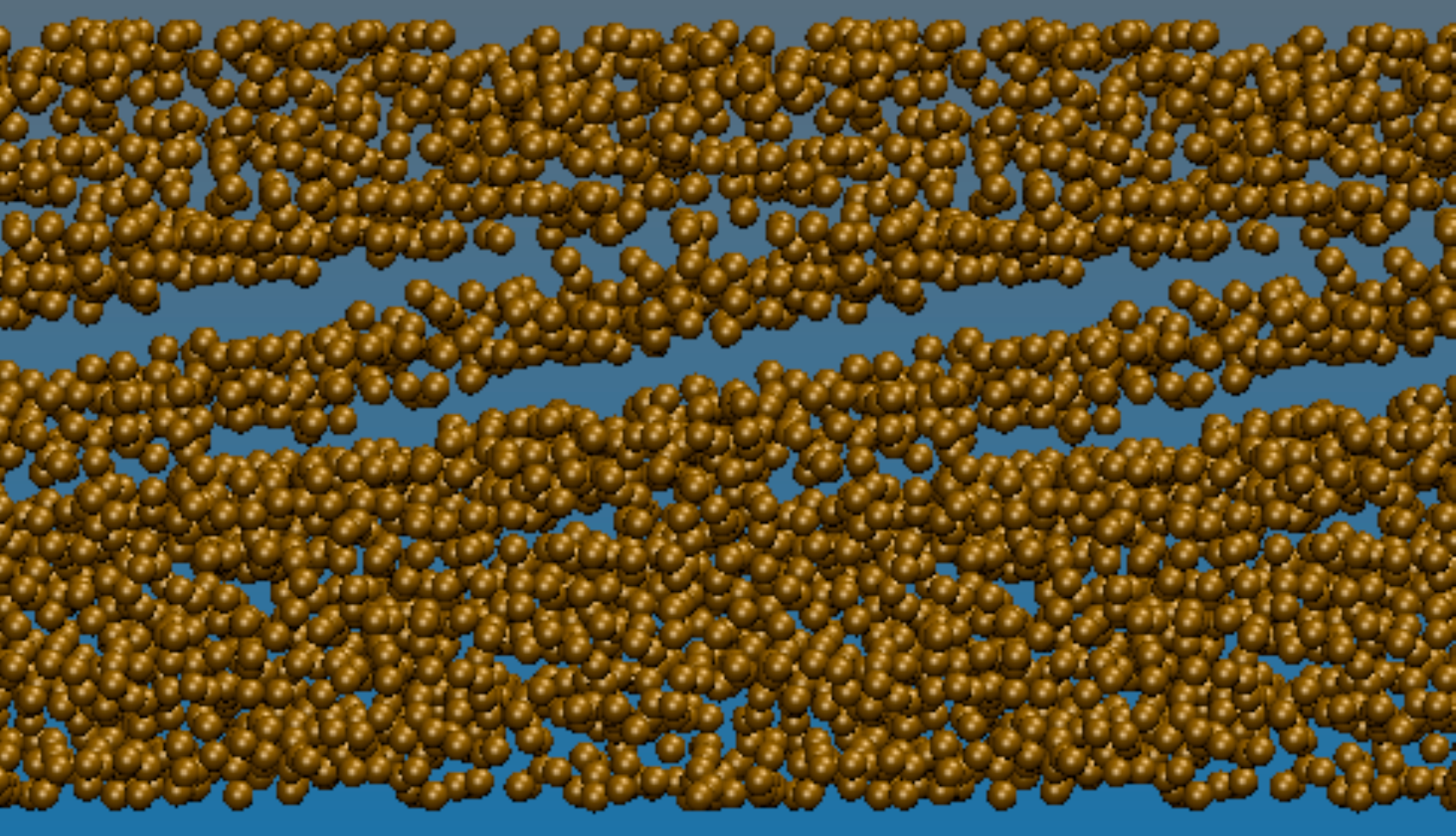}
\end{center}
\caption{ Snapshots of the simulation system B1 at different times during the stress overshooting experiment { at a constant shear rate of 100 $\tau^{-1}$.}
From top to bottom, the snapshots correspond to 0.5 $t_0$, $t_0$, 2.5 $t_0$, 5 $t_0$, 25 $t_0$ and 50 $t_0$, where $t_0$ is the time of the maximum of the stress. 
}
\end{figure*}

Viscoelastic materials like gels are typically characterized by the frequency dependent storage ($G'$) and loss ($G''$) moduli, that account for the elastic and viscous response, respectively. Experimentally, the viscoelastic moduli are obtained through the application of small amplitude oscillatory shear (SAOS) strain to the material and analyzing its stress response. For a pure elastic solid, the stress response will be in phase with the applied strain whereas the stress response of a pure viscous material will be 90$^\circ$ out of phase. Phase angles varying from 0 to 90$^\circ$ are characteristic of viscoelastic fluids. A similar approach was used to calculate the rheological properties of the simulated gels. As reported in our previous work\cite{Santos:2010aa}, a small amplitude oscillatory shear was applied to the simulation box and the resulting the stress tensor was used to calculate the viscoelastic moduli. Figure 2 shows the calculated $G'$ and $G''$ of three simulated systems (A1, B1 and B2), at a fixed frequency $\omega= 10^5 \tau^{-1}$, as a function of time during the gelation process. { At this frequency all the simulated systems display a gel like behavior, see Figure \ref{moduli2} below, with $G' > G''$.} For system A1, a sharp increase in the storage modules is observed at intermediate times ($t/\tau \sim 50$), concomitant to the sharp decrease in the potential energy shown in Figure 1. Systems B1 and  B2, which have a similar effective range but a lower temperature, show a smaller variation of the viscoelastic moduli than system A1, for $t/\tau  > 2$. The reasons is that for the lower temperature the gel develops very fast, and the kinetic of the system becomes very slow at earlier times as discussed above. Of the two systems at lower temperature, the one with shorter effective range (B1) forms a stronger gel. These results indicate that the strength of the gel, which is related to the storage modulus, is given by the overall structure of the percolated network and that the minor accommodations of the particles occurring at $t/\tau > 50$ do not significantly affect the macroproperties of the gel. The storage modulus is greater than the loss modulus showing the predominant solid characteristics of these gels at $\omega=10^5 \tau^{-1}$.

\begin{figure}[!b]
\begin{center}
\includegraphics[width=7.0cm]{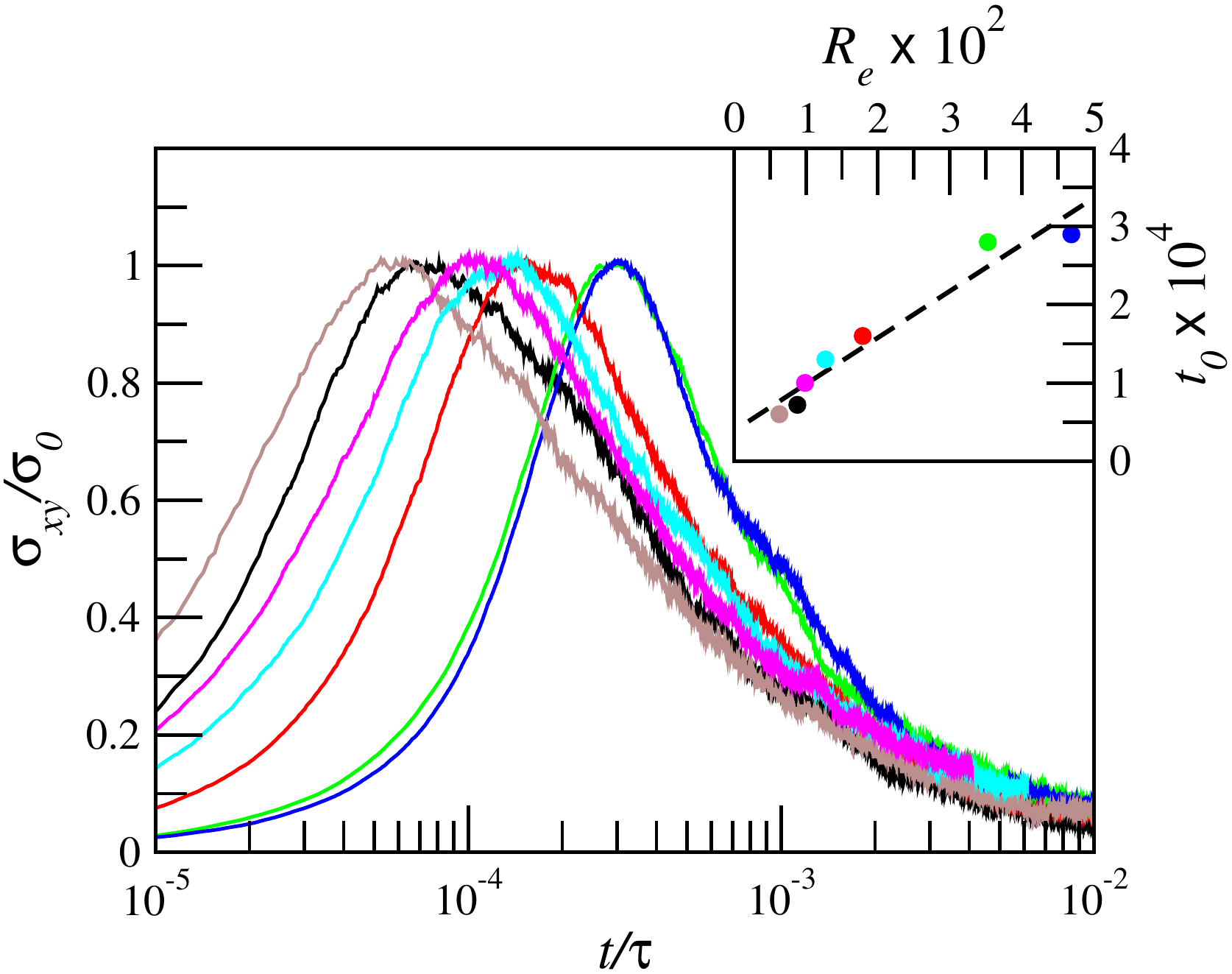}
\end{center}
\caption{Normalized stress overshoot for all systems undergoing a shear rate of $800 \tau^{-1}$. The insert shows the time of the maximum overshoot as a function of the effective range $R_e$, and the best linear fit is displayed by a dashed line. Colors as in Figure \ref{epot}}\label{allshoots}
\end{figure}


Alternatively to oscillatory techniques, transient methods may also be applied to study viscoelastic materials including start-up flow tests. In particular during this type of experimental characterization, viscoelastic systems often exhibit a distinct transient response known as stress overshoot. When shear is imposed to a viscoelastic fluid, the stress response of the material displays a peak (overshoot) before reaching a steady state value. This phenomenon is commonly observed in some polymeric solutions and melts, liquid crystalline polymers, nanofiber composites, foams, glass fiber suspensions and others.\cite{Mather:2000aa,Eberle:2008aa,Padding:2010aa,Kagarise:2010aa,Rajaram:2011aa} Start-up shear tests were also performed using Brownian dynamics simulations by applying a constant shear { rate} to the simulation box. { All these tests were performed on the structure obtained at $t=1000 \tau$.} We show in Figure 3 the stress response of the system B1 ($R_0=19$, $\sigma=1$ and $T^*=0.12$) for different shear rates. This is a typical case, all the simulated systems display a similar behavior. As the material starts to undergo a deformation, it behaves as an elastic solid with a sharp increase in stress. If the shear rate is sufficiently high, a maximum value is reached followed by a decay to the steady state value. For slow shear rates, the maximum is absent and the approach to the steady state value is from below. This behavior is observed experimentally and is related to stress relaxation. Low shear rates give the system time to relax and relieve the stress by structural rearrangements. On the other hand, high shear rates lead to a higher maximum stress since there is not enough time to rearrangements and relaxations of molecules/particles.\cite{Whittle:1997aa}. In real colloidal systems, the display of stress overshoot has been related to shear induced network fragmentation and particle orientation.\cite{Hoekstra:2003aa,Mohraz:2005aa,Letwimolnun:2007aa} Our systems undergo this kind of process, as can be observed by the sequence of snapshots, displayed in Figure 4. The snapshots reflect the state of the system before, at and after reaching the maximum stress, which occurs at $t_0$. It can be seen from the first three snapshots that the overshoot occurs with the system practically unperturbed by the applied strain, suggesting that it is due to the elastic response of the percolated network. Namely, the stress overshoot is the manifestation of a small perturbation to gel structure that induce separation between the particles and consequently the development of restoring forces by the inter particle potential. A clear deformation is observed for longer times ($> 5 t_0$) as the system develops layers parallel to the direction of the applied strain.

In order to further understand the origin of the stress overshoot, we show in Figure \ref{allshoots} the stress vs. time curves for all the simulated systems, for an applied shear rate of $800 \tau^{-1}$. In order to better visualize the results, we have normalized the stress by its maximum value for each system. From the figure emerges a clear linear relation between the effective range $R_e$ and the time at which the maximum stress occurs, $t_0$. This relation is displayed in the inset of Figure \ref{allshoots}. Intermolecular potentials with a longer effective range of attraction develop the maximum stress at later times. This result support our interpretation of the overshoot as simply the restoring force exerted between the particles of a slightly deformed, but structurally intact, percolated cluster. The restoring force is able to act for a longer time for those systems for which the effective attractions have a longer range. The linear relation between $t_0$ and $R_e$ suggests that a simple experimental method can be developed to estimate the value of the effective attractive range of real systems, providing a way to get insights on the inter-particle interactions from simple rheological experiments of macroscopic gels.

\begin{figure}[!t]
\begin{center}
\includegraphics[width=7.0cm]{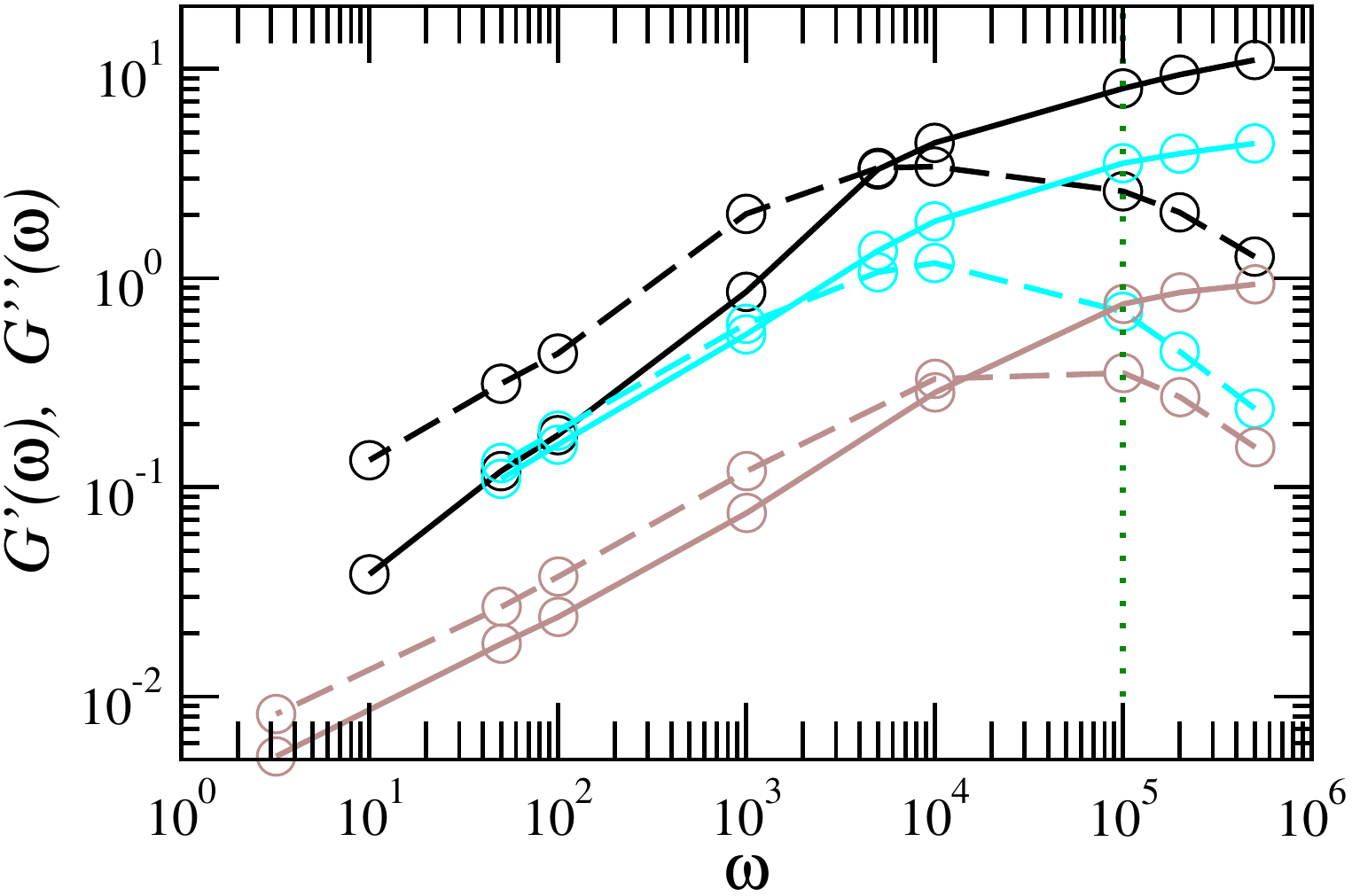}
\end{center}
\caption{Storage ($G'$, solid lines) and loss ($G''$, dashed lines) moduli for systems A1, B1 and B2, as a function of shear frequency. The green dotted line indicates the frequency of the calculations of Figure 2}\label{moduli2}
\end{figure}

\begin{figure}[!t]
\begin{center}
\includegraphics[width=7.0cm]{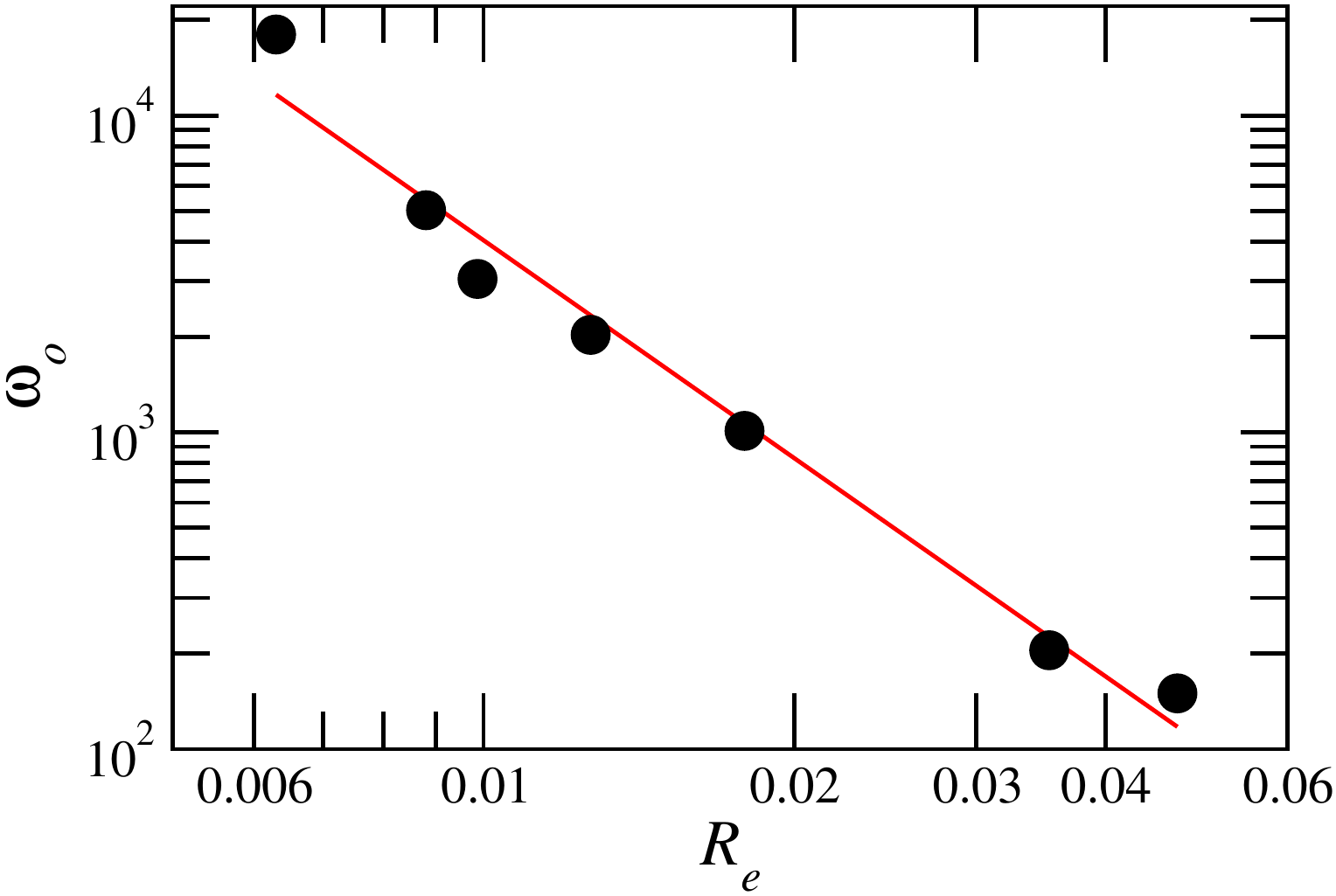}
\end{center}
\caption{Relation between the crossover frequency and the effective attractive range. The straight line is a power law that best fits the data, $\omega_0= 0.11\times R_e^{-2.3}$
}\label{cross}
\end{figure}

The transient gels formed after $t/\tau=1000$ were further investigated and their viscoelastic properties were related to the corresponding effective range $R_e$. The simulated frequency sweep tests performed in the linear viscoelastic region showed that for all studied systems the storage ($G'$) and loss ($G''$) moduli of the colloidal gel were dependent on the frequency applied. Figure \ref{moduli2} shows the typical viscoelastic curves obtained for three simulated conditions with different effective range $R_e$.  At very low frequencies the system should follows the characteristic limit for viscous fluids, namely $G' \propto \omega^2$ and $G'' \propto \omega$, however this limit is not reached in the present work due to the enormous computational cost of low frequency oscillatory shear simulations. At low frequencies, the loss modulus is the predominant response of the material. As the frequency increases, both $G'$ and $G''$ increase until they reach the same magnitude at the cross over frequency, $\omega_0$. At high frequencies, the gel displays solid like characteristics, reflected by a large storage modulus. As in the case of the start-up shear tests, the effective attractive range characterizes a well defined property that is accessible from straightforward experiments. The crossover frequency, plotted in Figure \ref{cross}, decreases as the attractive effective range increases. Moreover, the results are well represented by a power law relation. The negative correlation between $\omega_0$ and $R_e$ appears to be counterintuitive. On the one hand, it has been well established that short range potentials form gels with a much longer life than gels formed by short range potentials. On the other hand, the relation between the cross over frequency and the effective range suggests a gel like material for a wider frequency range is obtained for the inter particle attractions with long range. This behavior has been discussed by the early work of Lodge and Heyes\cite{Lodge:1999aa} in their study of $2n-n$ Lennard-Jones systems and our previous work\cite{Santos:2010aa}, and the two apparently contradictory statements can be reconciled: Provided that the system has a short enough attractive range to develop a relatively long lived gel, the characteristics of this system shows a higher elastic behavior for the longer effective attractions.

\section{Conclusions}

We have performed a series of Brownian dynamics simulations on model colloidal gels, varying the size of the particles, temperature of the system and magnitude the attractive tail of the particle interaction. The gels are formed by destabilizing a high energy dispersion by instantaneous quenching of the system to the target temperature. We observe the formation of a percolated cluster in a relatively short time for all the studied systems. The long time dynamics is mainly dominated by the system temperature, with the lower temperature undergoing a kinetic slowdown.

We present for the first time an analysis of the viscoelastic response of colloidal gels in terms of an effective attractive range of continuous inter-particle potentials. The theory of Noro and Frenkel for corresponding states was applied, and in particular the definition of the effective attractive range, $R_e$, to analyze the viscoelastic properties of the simulated colloidal gels. It was found that $R_e$ is a key parameter determining the time $t_0$ to reach the maximum stress in start-up shear and the crossover frequency $\omega_0$ in small oscillatory frequency sweep simulations: $t_0$ is a decreasing linear function of $R_e$ whereas $\omega_0$ follows a power law. These two relationships, in principle, could be used to obtain a quantitative estimation of the effective attractive range of colloidal particles interaction through straightforward standard rheological methods.





\section{Acknowledgments}
This work was supported by the U.S. Army Research Office under the Multi-University Research Initiative (MURI) grant number W911NF-08-1-0171.

\footnotesize{
\bibliographystyle{rsc} 
\providecommand*{\mcitethebibliography}{\thebibliography}
\csname @ifundefined\endcsname{endmcitethebibliography}
{\let\endmcitethebibliography\endthebibliography}{}

}

\end{document}